# Quantitative Indices for Improving Metro Load Curve, Using Distributed Generation

Masoud Behbahani
*Department of Power Engineering, Railway Industries Group*
*TAM Iran Khodro Co.*
*Tehran, Iran*
m.behbahani@tam.co.ir

Alireza Fereidunian
*Faculty of Electrical Engineering,*
*K. N. Toosi university of Technology*
*Tehran, Iran*
fereidunian@eetd.kntu.ac.ir

*Abstract* - **This paper promises the idea of using DG (Distributed Generation) to improve the Metro load curve. Public transportation systems are often based on gasoline and diesel. However, with the gradual development in usage of the Metro and monorail, a new load with heavy demand, inappropriate load curve and middle LF (Load factor) is added to the electricity grid. In addition to supply problem of this massive consumer, the Metro load curve is another problem, which has a relatively low LF. Furthermore, Metro load peak hours coincide with the peaks of national grid. Improvement of the load curve is well-known in electrical engineering literature, which depending on the type of load curve, offers general recommendations in three approaches; DSM (Demand Side Management), DS (Distributed Storage) and DG. In this paper, to achieve quantitative indices of improvement for Metro load curve using DG, firstly based on the analysis of volume and consumption pattern of the main loads in Metro, the typical load curve has been extracted. Using this curve, the result of using DG is shown by quantitative parameters which represent the significant improvement in load curve. These parameters can be used to calculate economic indicators such as initial cost and ROI (Return of Investment).**

I. INTRODUCTION

Population growth and other problems like traffic and pollution has led to an increasing tendency to use public transport systems. Among these, electrically fed systems like Metro, Monorail and Trolley buses have received more attention, due to their low $CO_2$ emission. In most countries, statistics show increased usage of these systems. In recent decade, the Tehran Metro has become an integral part of the public transportation system in the Tehran metropolitan area. Further to development phases of the Tehran Metro, other large cities have started Metro and monorail projects, so that the "Energy Balance" Report of the Iranian Ministry of Energy indicates 22.6% growth in power consumption of the Metro during the fourth National Development Plan for the period of 2005 to 2009 [1].

Fig. 1 shows the actual energy consumption of Metro system from 2004 to 2010 and its forecast from 2009 to 2016 [1] [2]. This prediction has considered the development phases of Tehran Metro and Metro projects in Mashhad, Tabriz, Ahwaz, Esfahan, Karaj and Shiraz. Meanwhile, new projects like Qom Monorail, Qom Metro and Kermanshah Monorail have been started recently.

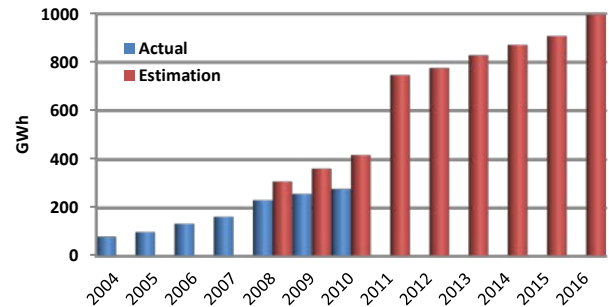

Fig.1: Power consumption in Urban Railway of Iran

Because of the high energy demand of Metro system, its supply through urban MV network is almost impossible, thus it needs independent substations and transmission lines, which requires enormous investment.

On the other hand, traditionally, the major loads of electricity grid are divided into five types: household, industrial, commercial, agricultural and public loads. Each type possesses known load characteristics and the national grid load curve is formed with a weighted combination of them. In Metro, the bulk load is directly proportional to the number of passengers. The morning and evening rush hours produce two peaks in trains' traffic and the Metro load curve. Fig. 2 shows the total number of passengers in and out of London stations [3]; while Fig. 3, shows the dynamics of a typical weekly hourly load curve in megawatt hours of Italian railway [4].

Therefore, in addition to high demand of urban railway, it's inappropriate load curve imposes problems to the grid. Metro load curve shows that the grid shall supply the load that coincided to domestic and commercial peaks. This increases the peaks which build considerable difficulties [16]. The Electrical Engineering literature possesses well-known principles to improve the load curve, suggesting methods, related to load curve type [5] [6].

This paper promises the idea of using DG (Distribute Generation) to improve the Metro load curve. First, a short explanation is given on the existing technologies in railway industry, which positively improve the load curve. Then, after developing a typical Metro load curve based on a real time schedule, the improvement results of using DG have been investigated by providing quantitative parameters.

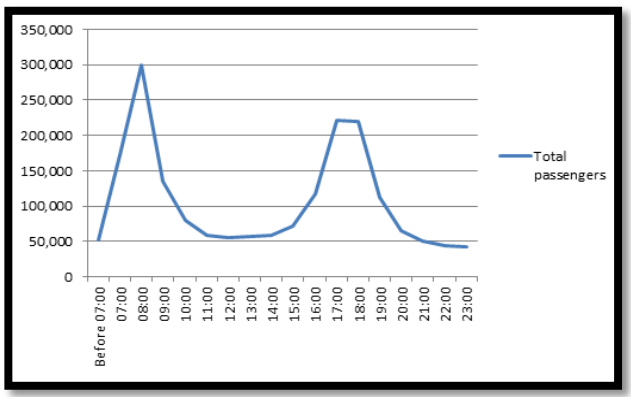

Fig. 2: Passenger number in London Metro Station

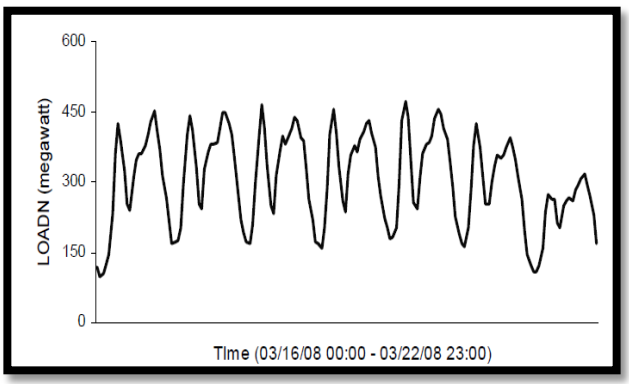

Fig. 3: Weekly hourly load curve of the Italian railways

II. LOAD CURVE IMPROVEMENT APPROACHES

A. DSM (Demand Side Management)

From the consumer point of view, DSM is defined as the following methods:

1) Peak Clipping
2) Valley Filling
3) Load Shifting
4) Strategic Conservation
5) Strategic Load Growth
6) Flexible Load Shape

Peak Clipping means reduction of dedicated loads for the safety of passengers. Other methods except the strategic conservation are dependent on actual working hours of offices and markets, thus they require changing the passenger traffic, which is out of scope of the urban railway organization or national grid.

Strategic conservation or simply, energy saving is received more attention recently due to high costs, limited resources and environmental pollution of fossil energy. This approach presents new solutions according to technology advancement. Energy saving can be considered the best choice, since it solves the problem at the point of consumption and follows considerable saving duo to reduction of generation and transmission losses.

Metro loads are supplied by DC voltage and AC low voltage. The first is provided by TPS and used for traction; while the second is provided by LPS and used for all other Metro consumers such as lighting, receptacles, elevator, escalator, water pump, HVAC and etc. Strategic conservation for LPS consumers will be realized by traditional methods for energy saving in building and substations. For traction, this requires specific technologies of railway industry, which are based on two pillars, movement management and efficiency of fleet.

Fleet Movement management includes methods such as automation, intelligent time table and speed control in both manual and automatic modes, that can be referenced as example in [8] and [9]. Fleet Efficiency increasing is combined from different technologies such as replacing DC traction motor with Induction motor or PMSM, efficiency of auxiliary equipment such as lighting and HVAC, efficiency of traction motors drive, improve thermal insulation, low weight and aerodynamic fleet [10-11].

B. DS (Distributed Storage)

Often the extra kinetic energy of fleet is driven out as heat, using mechanical and electrical brakes. Regeneration of this energy and way back to the electricity grid or charge equipment such as batteries, super-capacitors, Ultra-Capacitors or special flywheels and then discharged them in accelerating time are proposed technologies for DS. In this context, there are numerous references that [12] and [13] provide appropriate conclusion.

C. DG (Distributed Generation)

Many advantages are stated for DG, that the base of all of them is energy production in nearer distance to load. Example advantages can be named as reducing received energy from national grid, reducing the effects of network disturbances, reducing the demand cost, reducing transmission losses and reduced investment in transmission lines. Due to amount and concentration of the industrial loads, DG is mainly focused on industries. This also receives appropriate incentives from the environmental and governmental organizations like the Iran Energy Efficiency Organization (IEEO -SABA) recently.

Between all of the subscribers of Tehran Regional Electricity Company (TREC), few industries operate emergency diesel generators. Among those, Pars Khodro Co. and IKCO (Iran Khodro) are the industries which support the grid to reduce peak power. This produces a significant share of the peak decline with providing 5.52MW demand and 2054MWh energy [14]. A similar study in Mazandaran industries has shown that emergency generators have a potential to reduce 24MW from peak power, thus a 2.1% increase in LF (Load Factor) [15]. Because of low density and low demand in other traditional users, using DG has lower priority in technical terms. It also receives less attention from economic perspective due to considered tariffs.

In Rail industry, unlike DSM and DS, there are few DG application references or case studies, while the Metro demand is about a few tens of MW and are comparable with heavy industries [14]. Due to essential safety factors in Urban Railway, TPS and LPS substations are fed by one or two independent rings. Thus although Metro is runs around the city like residential and commercial consumers, feeding by DG is technically feasible, since it is supplied by an exclusive network,

### III. EXTRACTION OF THE METRO LOAD CURVE

Load curves can be extracted by measuring the energy of outgoing feeders at different times. In this paper, the typical load curve is extracted based on the behaviour and quantity of common loads in the Metro

A. Traction Power Substation

TPS is supplied by medium voltage level, converting MV to DC voltage and supplying required energy for traction. Obviously the TPS load curve is proportional to the number of running trains, and thus is relative to the number of passengers. Accordingly, the Fig. 4 is used as typical load curve of a traction system, which shows the system energy consumption as a function of number of passengers.

According to train's timetables in the Tehran metro, the number of trains at intervals of a working day is plotted in Fig. 5. The curve is similar to Fig. 2-4 shows two peaks in the morning and in the evening.

B. Light and Power Substation (LPS)

LPSs are supplied by medium voltage level, converting MV to LV and supplying all the other consumers except the traction. An overview of the typical load of LPS indicates that some loads are mainly proportional to the number of passengers and the rest, are independent. In the Metro, tunnel ventilation, lifts and escalators are examples of applications that depended in part on the number of passengers; while lighting, electricity outlets, signalling, telecommunication, control and HVAC are independent consumptions. According to relative contribution of these two groups, the typical load curve of LPS feeders is extracted in Fig. 6.

Experimentally, LPS power requirements is between 50% to 70% of the TPS, so assuming a 60% combined ratio of these two loads is shown in Fig. 7, the LF is calculated as 0.53 according to (1).

$$LF = \text{Average Load/Maximum Demand} \qquad (1)$$

The calculated LF is based on the assumptions considered for simplification of calculation; in a practical project measures taken outlet feeders can be a criterion to determine the load curve. The low LF of Metro system means a significant investment should match to capacity during rush hours, that is unused during rest of time. In other words, "The problem with running a railway is that you have to invest in loads of rolling stock to meet rush hour demand, and then it sits there idle most of the day" as clearly stated by [3].

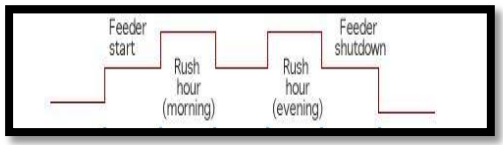

Fig. 4: Typical Load Curve of a TPS

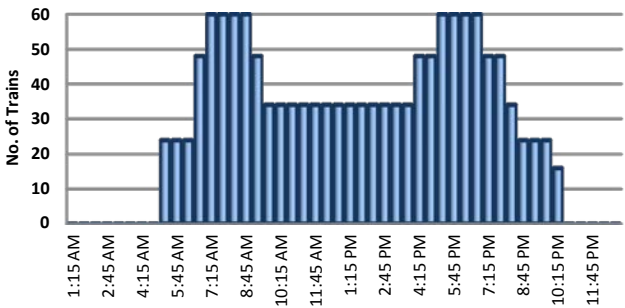

Fig. 5: No of Trains in a Metro

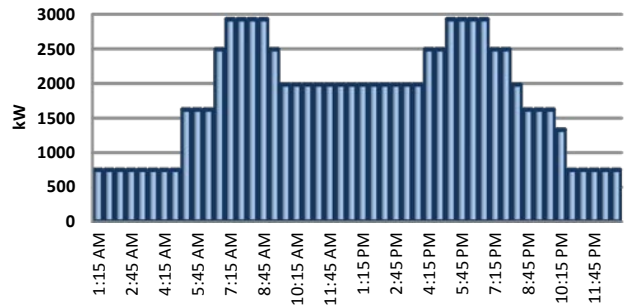

Fig. 6: LPS Load Curve

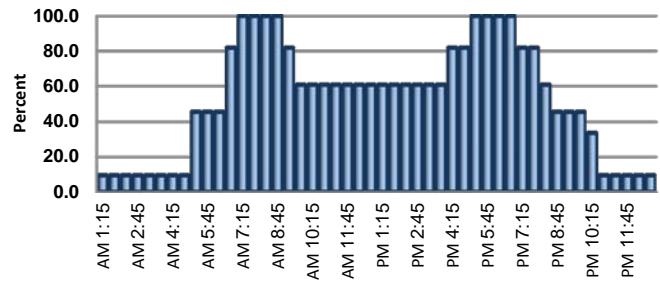

Fig. 7: Metro Load Curve

## IV. Deploying DG in Metro

### A. Case Study

Compared to the industry, the Metro owns a worse load curve. Therefore, the use of DG in Metro power system has more benefits rather than the other ones, due to providing morning and evening peaks. All of these factors can reduce ROI in comparison with other application and produce more benefits for national grid and urban Rail Organization. This paper proposes supplying Metro using DG in parallel to national Grid, in which the capacity of DG is defined by (2).

$$P_{DG} = P_{Peak} - P_{base} \qquad (2)$$

Where, $P_{peak}$ is the maximum value of the load curve and $P_{base}$ is the average value of the load curve in the interval between 9:30 am to 4 pm. $P_{DG}$, the capacity of DG is prepared by one or more diesel generators. For example, in load curve of Fig. 7, $P_{base}$ equals to 60% of $P_{peak}$, thus $P_{DG}$ can be calculated as:

$$P_{DG} = P_{Peak} - 0.6 * P_{peak} = 0.4 * P_{peak} \qquad (3)$$

Therefore, using DG with rating power equal to 40% of peak load, the load curve can be improved immensely. Both load curves are compared in Fig. 8.

### B. Discussions

According to the previous section, the suggested DG works from 7 to 9 am and 5 to 7 pm, to coincide with morning and evening peaks, which results the following extra benefits:

- ✓ A 40% reduction in electricity demand
- ✓ A 64% reduction in transmission losses at peak times [16].
- ✓ An improvement of the Metro LF from 53% to 73%
- ✓ A possibility of dismantling the dedicated emergency diesel generators
- ✓ A possibility of reducing the required battery capacity

Due to the high safety requirements in urban railway, many Metro loads require backup power systems such as emergency diesel generators and UPS. Using DG can eliminate the emergency diesel generator and also reduce support time of UPS i.e. reducing the battery capacity, that cost savings will be significant.

The grid uses gas turbine plants to meet peak load, thus the cost of produced energy does not change considerably. However, considerable cost reduction can be calculated according to $P_{DG}$ in transmission process.

Technologies such as energy saving and DS in Rolling Stock or automation are expensive and the monopoly of the international companies, which are selected based on the compromise between cost and performance on purchasing phase. In contrast, energy saving in LV consumers, speed control, intelligent timetable and DG are technologies which can be followed in parallel of available science, technology and facilities and relying on domestic capabilities.

## V. Conclusions

Extracting a typical load curve, allows the quantification of economic indicators of using DG for improving the Metro load curve. Compared to the other network consumers, Metro system has a better techno-economical justification for using DG, due to its type of Load Curve, dedicated network and the heavy demand. This techno-economical assessment should be evaluated for each Metro or Monorail based on its actual load curve. The calculated first investment and ROI value can show the extra benefits of the presented methodology in comparison with full grid supplied urban railway. This approach can prevent from loss of assets, in a national broader of view.

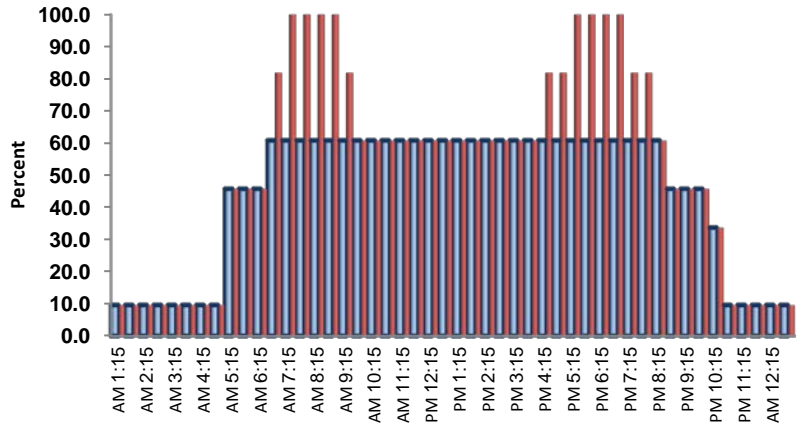

Fig. 8: Metro load curve before vs. after using DG